%% file: manuscript.tex
\documentclass[aip, graphicx, tightenlines, 12pt, reprint, floatfix, superscriptaddress]{revtex4-1}
\usepackage[utf8]{inputenc}

\usepackage{graphicx}
\usepackage{amsmath}
\usepackage{amssymb}
\usepackage[hidelinks]{hyperref}
\usepackage{multirow}
\usepackage{soul}
\usepackage{xcolor}

\usepackage[font=small,labelfont=bf]{caption} 

\usepackage[version=3]{mhchem} 

\input{commands}

\begin{document}

\title{The fundamental drivers of electrochemical barriers}
\author{Xi Chen}
\affiliation{School of Engineering, Brown University, Providence, Rhode Island, 02912, USA}
\author{Georg Kastlunger}
\affiliation{Department of Physics, Technical University of Denmark, DK-2800, Kongens Lyngby, Denmark}
\author{Andrew A. Peterson}
\email{andrew\_peterson@brown.edu (corresponding author)}
\affiliation{School of Engineering, Brown University, Providence, Rhode Island, 02912, USA}

\begin{abstract}
\noindent
We find that ion creation/destruction dominates the behavior of electrochemical reaction barriers, through grand-canonical electronic structure calculations of proton-deposition on transition metal surfaces.
We show that barriers respond to potential in a nonlinear manner and trace this to the continuous degree of electron transfer as an ion is created or destroyed.
This explains both Marcus-like curvature and Hammond-like shifts.
Across materials, we find the barrier energy to be driven primarily by the charge presented on the surface, which in turn is dictated by the native work function, a fundamentally different driving force than non-electrochemical systems.
\end{abstract}

\maketitle

Electrochemistry is a linchpin technology in the transition from fossil fuels, providing short- and long-duration storage as well as long-distance movement of intermittent electricity~\cite{Yang2011,Miao2021}, vehicle power~\cite{Hoekstra2019,Cunanan2021}, and pathways to the defossilization of countless industries, including synthetic fuels\cite{Dietenberger2007}, cement~\cite{Ellis2019}, fertilizers~\cite{Westhead2021}, steel~\cite{Lopes2022}, aluminum~\cite{Pawlek2016}, and (non-fossil) petrochemicals~\cite{IEA2018}, either directly or through hydrogen intermediates.
While electrochemistry has a rich history in analytical chemistry, our understanding of the controlling reactions at the level of electronic structure calculations is still emerging.
Sophisticated electronic-structure approaches have recently gained popularity due to the availability of methods that allow the simulation of reactions at constant applied potential, which naturally occur in the electronically grand-canonical ensemble.~\cite{Alavi_JCP_2001, Sugino_PRB_2006, Neurock_2006, Jinnouchi2008, Letchworth-Weaver2012, Head-Gordon_2016,Sundararaman2017, Kastlunger2018, Bouzid2018, Melander2019, Hormann2020, Hormann2021,Lindgren2022}
Advanced simulation methods mimic physical potentiostats to maintain a constant electrical potential, by varying the (fractional) number of electrons while allowing for a compensating, screened countercharge as well as a field-free region providing an absolute reference for the electrochemical potential of electrons.

Reaction barriers dictate the rates of elementary reactions, and therefore are crucial to understanding electrochemical kinetics.
In this work, we employ the solvated jellium method~\cite{Kastlunger2018} to understand the basic driving forces behind electrochemical reaction barriers, uncovering their similarities to as well as highly notable dissimilarities from conventional reactions at surfaces, such as those in heterogeneous catalysis.
Fundamental electrochemical reactions often involve the creation or destruction of an ion---such as the conversion of \ce{H+} to adsorbed hydrogen (and ultimately \ce{H2} gas), the conversion of graphene-adsorbed lithium to \ce{Li+} in solution, or the conversion of \ce{CO2} to adsorbed \ce{COO-}.
Such reactions are inherently coupled to the movement of electrons to reaction sites, and we will show that the barriers for such reactions are strongly affected by this electron transfer as well as electrostatic interactions---both of which vary along the reaction path---making an understanding distinct from those of conventional surface reactions.
We focus on trends in the simplest such reaction: the proton-deposition reaction (\ce{H+ + $x$ e- -> H$*$}, where $*$ is the electrode surface and the amount of electron transfer $x$ is determined \textit{a posteriori}~\cite{Fang2014,Ge2020,Hormann2020,Lindgren2022}), but we expect the concepts and trends we develop to hold across ion-creating or -destroying reactions at electrochemical interfaces.
For reasons of generality, we neglect tunneling or vibronic effects, which are treated by others.~\cite{Lam2020,Melander2020}

\begin{figure*}
\includegraphics[width=1.0\textwidth]{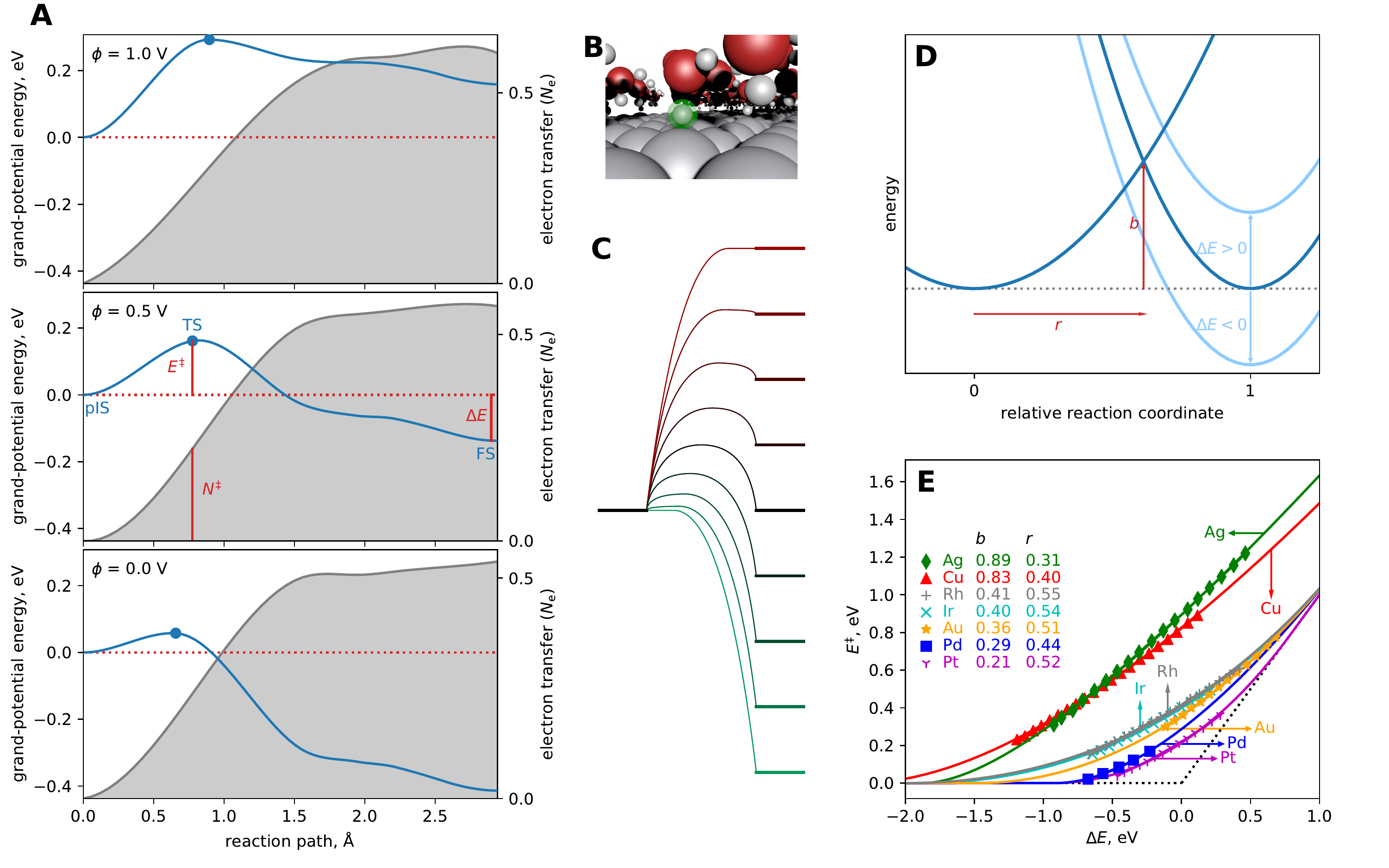}
\caption{
\textbf{Potential-dependent behavior of reaction barriers.}
\textbf{A}:~Reaction barriers and electron transfer on Pt(111) at three fixed potentials, versus standard hydrogen electrode (\vshe).
Constant-potential reaction paths were found by pairing the solvated jellium method~\cite{Kastlunger2018} to calculate grand-canonical forces with the DyNEB method~\cite{doi:10.1021/acs.jctc.9b00633} to search for the saddle point.
Potentials are referenced assuming SHE is 4.44~V versus vacuum, as suggested by Trasatti~\cite{Trasatti1986}.
\textbf{B}:~Atomic figure of the transition state on Pt(111) at 0.5 \vshe.
\textbf{C}:~Schematic of the Hammond--Leffler postulate, which can be interpreted to mean that strongly downhill reactions have barriers similar in structure and energy to the initial state, and vice versa.
\textbf{D}:~Two-parameter model for curved barrier scaling.
\textbf{E}:~Barrier heights as function of energy change for several transition metals.
\label{fig:Pt-barriers}}
\end{figure*}

In Figure~\ref{fig:Pt-barriers}\textbf{A} we show converged calculations of a coupled proton--electron transfer to a Pt (111) surface, with all solvent and surface degrees of freedom relaxed, at three different applied potentials \potential.
At each potential, we plot the grand-potential energy versus the reaction coordinate, which starts from the pseudo-initial state (pIS) and proceeds through the transition state (TS) to the final state (FS).
We use the term pIS, rather than IS, as the pIS involves a solvated proton (\ce{H3O+}, \ce{H5O2+}, etc.) localized near and hybridized with the surface; in kinetic models the pIS should be properly referenced to a thermodynamic initial state, as discussed in-depth elsewhere.~\cite{Chen2018, Lindgren2022}
Here, we focus solely on the elementary process.
A still image of the TS at 0.5 \vshe\ is shown in Figure~\ref{fig:Pt-barriers}\textbf{B}.
At each potential, the charge transfer---that is, the number of electrons that must be injected by the ``computational potentiostat''---is also plotted.
The total electron transfer is significantly less than one, a consequence of the prehybridized pIS.\cite{Chen2018}
The degree of charge transfer at the barrier \barrierelectrons\ will become an important quantity in the subsequent analysis.

Several overall trends are apparent: as the potential becomes more negative, the reaction becomes more downhill (exothermic), the barrier grows smaller, and the position of the transition state shifts earlier, towards the pIS.
These latter observations are consistent with the Hammond--Leffler postulate~\cite{Leffler1953,Hammond1955} (shown qualitatively and described for general reactions in Figure~\ref{fig:Pt-barriers}\textbf{C}), and later we will derive quantitatively why this behavior holds for electrochemical reactions.
We employ a simple model to correlate the barrier height \barrierheight\ to the energy change \reactionenergy, shown in Figure~\ref{fig:Pt-barriers}\textbf{D}.
In this two-parameter expansion of a previous model~\cite{Lindgren2020}, $b$ represents the barrier height at $\reactionenergy=0$, while $r$ indicates the degree of skew (\textit{i.e.}, asymmetry of parabola widths), with a symmetric response given by $r=\frac{1}{2}$.
(While this bears a superficial resemblance to Marcus theory, we employ it solely to match the limiting behavior of the Hammond--Leffler postulate.)
The full form and derivation of this model are shown in the SI.
In Figure~\ref{fig:Pt-barriers}\textbf{E}, we show the results of extensive grand-canonical barrier calculations across several transition metals and potentials; full results are shown in the SI.
From this figure, a curvature in the barrier height is clear, in contrast to predictions of the conventional BEP, transition-state, and linear free-energy scaling relations conventionally employed in catalysis and organic chemistry.~\cite{Evans1938,Bligaard2004,Wang2011a,Wang2011b,Sutton2012,Rosta2012,Hoffmann2022}

\begin{figure}
\includegraphics[width=1.0\columnwidth]{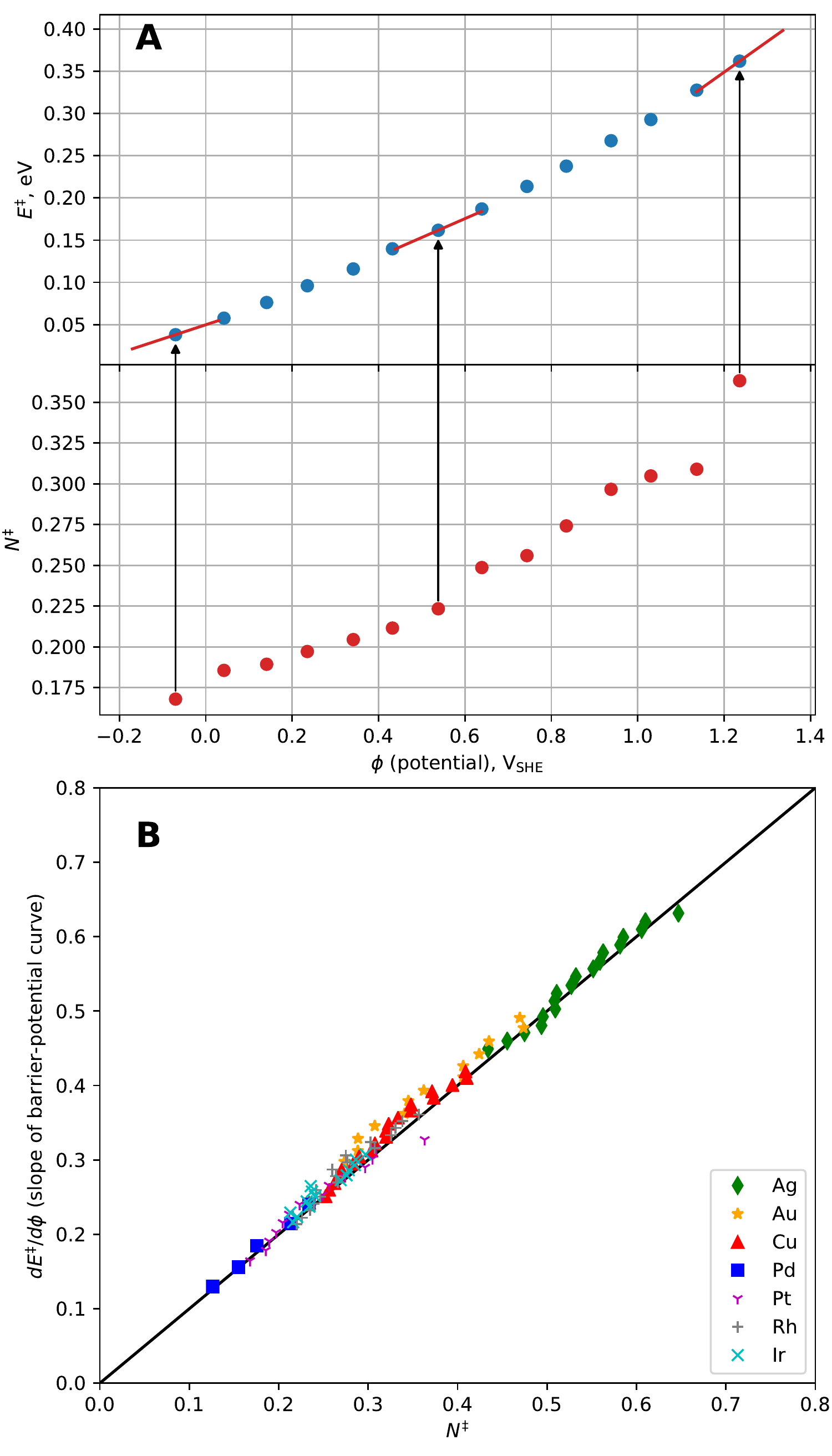}
\caption{
	\textbf{Electron transfer drives barrier's local potential response.}
\textbf{A}: Reaction barrier (\barrierheight) and electrons transferred at the barrier (\barrierelectrons) both plotted versus potential (\potential).
The slopes on the upper figure are those calculated from \barrierelectrons\ in the bottom figure.
\textbf{B}: The local slope of the barrier height versus potential ($\frac{d \barrierheight}{d\potential}$) curve (calculated by fitting the curve to a parabola and taking an analytical derivative) versus \barrierelectrons\ for each surface/potential combination.
\label{fig:barrier-electrons}}
\end{figure}

What drives the barrier to change with potential, and what is responsible for the curvature?
In the SI, we show that a simple thermodynamic relationship exists:

\begin{equation}\label{eq:barrier-change} \pd{\barrierheight}{\potential}{\pos} = e \barrierelectrons \end{equation}

\noindent
where \barrierelectrons\ indicates the amount of electron transfer at the barrier, and $e$ is the (positive) electronic charge.
This relationship is exact, in the limit of fixed atomic positions \pos.
Thus, to a first approximation, we can consider the barrier height to change in proportion to the amount of electron transfer at the barrier.
With grand-canonical methods, this quantity is unambiguous: it is simply the difference in the number of excess electrons at the barrier minus those at the reference initial state; this can be seen in the middle panel of Figure~\ref{fig:Pt-barriers}A.

This simplifies the analysis of free-energy diagrams: over small perturbations the barrier changes in an analogous fashion to how reaction endstates have long been assumed to change; that is, consistent with the widely-used computational hydrogen electrode\cite{Norskov2004,Peterson2010a,Chan2016,Hormann2020,Lindgren2020} approach, as $\Delta \barrierheight \approx e \barrierelectrons \Delta \phi$.
However, since \barrierelectrons\ itself is a function of potential \potential, this predicts the local slope only, with the ultimate expression exhibiting curvature. 
This differs from models based on capacitance,~\cite{Chan2016} which predict linear behavior.
In Figure~\ref{fig:barrier-electrons}, we show that this thermodynamic relation holds locally for Pt (111), and we also show that it holds locally for each metal surface we report in this study, across their full range of potentials.

\begin{figure}
\includegraphics[width=1.0\columnwidth]{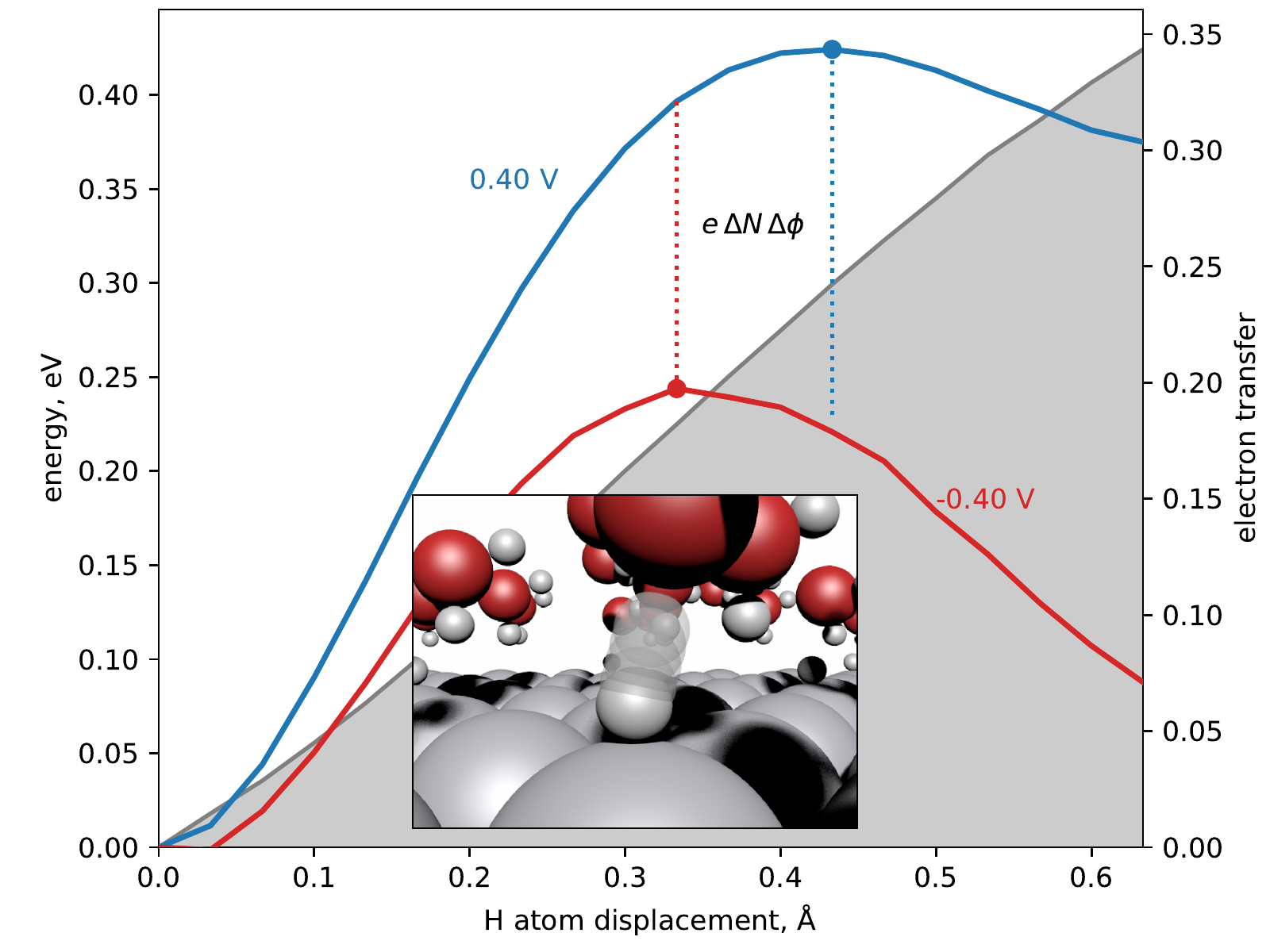}
\caption{
	\textbf{Electron transfer drives barrier movement.}
The degree of charge transfer along the reaction path drives the barrier to move earlier and become smaller as the potential is made more negative.
	The top (blue) curve shows the barrier of a fixed path at +0.40 \vshe, while the shaded curve shows the charge transfer at each image along the path.
	The vertical lines coming down show quantitative predictions of the energy change, at each image, based upon $e \, \Delta N \, \Delta \phi$.
	The bottom (red) curve shows the DFT-calculated barrier of the same fixed path at -0.40 \vshe.
\label{fig:moving-barrier}}
\end{figure}

We can use the relationship of equation~\eqref{eq:barrier-change}---which holds for any two fixed geometries---to analyze this behavior in more depth.
To do this, we contrive a simple one-dimensional system shown in Figure~\ref{fig:moving-barrier}, where the proton is constrained to move along a one-dimensional interpolation between the water layer and the surface.
Every other atom was fixed at its pIS position, to ensure only a single reaction path is possible regardless of potential.
The energy of each state along this path is shown in the top curve at 0.4 \vshe, and the peak of this path can be considered the barrier height \barrierheight\ for this system.
The shaded curve shows the corresponding electron transfer for this system.
From equation~\eqref{eq:barrier-change} we can predict that the energy of any state along this curve will be reduced by $e \Delta N \Delta \potential$, where $\Delta N$ is the electron-transfer relative to the start of the curve.
We quantitatively predict this change for the barrier and one earlier image in the figure, shown by the dashed vertical lines.
Since more electrons have transferred at the barrier, the barrier energy decreases more rapidly with potential, relative to the earlier state; this shifts the maximum earlier while the energy decreases.
This thus predicts how the barrier shifts in both magnitude and position as the potential is changed, and we see this prediction is a near-exact match to grand-canonical electronic structure calculations carried out at the lower potential of -0.4~\vshe.

Thus, the charge transfer along a reaction path, in combination with equation~\eqref{eq:barrier-change}, predicts and explains a multitude of related effects: the observed Hammond--Leffler behavior of barrier movement, the energetic shifts of barriers with potentials, and the curved Marcus-like behavior of barrier energies.

\begin{figure*}
\includegraphics[width=1.0\textwidth]{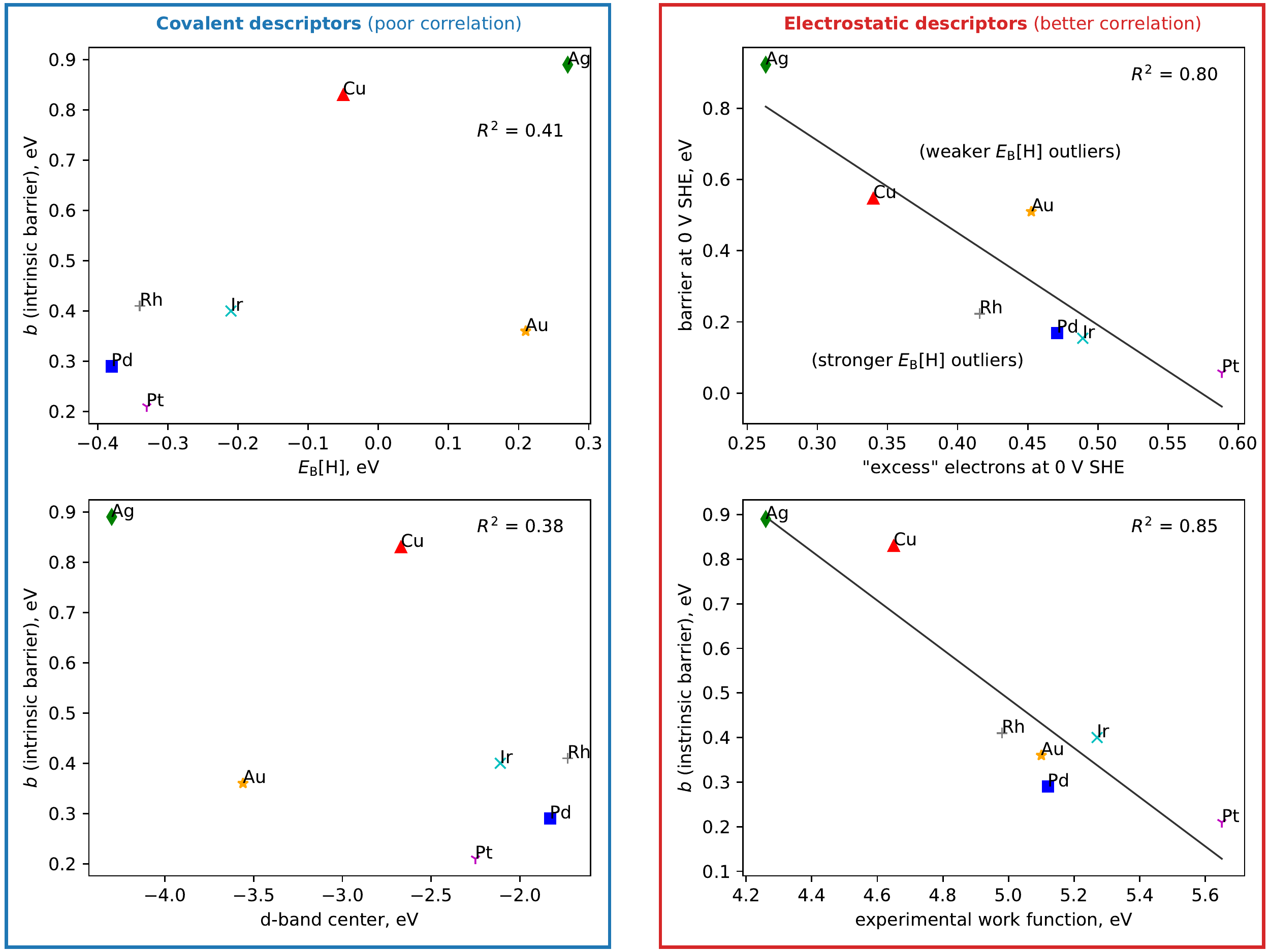}
\caption{\textbf{Correlation of barriers with various material properties.}
The barrier exhibits a strong correlation with electrostatic descriptors, and only a weak correlation to covalent descriptors.
\label{fig:material-correlations}}
\end{figure*}

We next turn our attention to trends among materials, attempting to understand if electrochemical barriers follow the general trends developed for catalytic systems.
These commonly used trends include correlations of the reaction barrier to the binding energies of reactants/products~\cite{Bligaard2004,Sutton2012}, and---for late transition metals---correlation of adsorption energies in general to the central moment of the material's d-band.~\cite{Hammer1995,Abild-Pedersen2007a}
In our system, the FS is a hydrogen bound to a metal surface, while the TS is a hydrogen loosely bound to the same surface, so we would na{\"i}vely expect the TS energy to correlate to the FS energy across metals.

In Figure~\ref{fig:Pt-barriers}\textbf{E} we showed the relationship between \barrierheight\ and \reactionenergy\ for a range of late transition metals.
We can use the parameter $b$ to make comparisons between materials, as it  corresponds to the barrier height with zero local driving force; we'll refer to this as the intrinsic barrier.
Interestingly, we see that Pt has an abnormally low intrinsic barrier, perhaps contributing to the many proposed reasons for platinum's excellence in catalyzing the hydrogen evolution reaction.
Conversely, we see relatively high intrinsic barriers for Ag and Cu.

In Figure~\ref{fig:material-correlations}, we first attempt to correlate the intrinsic barriers to the materials' hydrogen binding energies and d-band centers.
We see no strong correlation of the intrinsic barrier to either of these quantities.
This suggests a different factor may be driving the barrier energetics.

A key differentiating feature of electrochemistry is that reactions typically involve creating or destroying an ion, and thus the ionic nature of the transition state may play a crucial role, which we next explore.
Each material requires a different amount of electronic charging to reach a particular voltage; therefore, at any particular voltage each material expresses a different surface charge.
In the top-right panel of Figure~\ref{fig:material-correlations} we focus on the 0 \vshe\ condition, and plot the barrier versus the charge presented on each electrode; we now observe a strong correlation---suggesting an electrostatic driving force.
Interestingly, we see that the hydrogen binding energy trends predict the deviation from the correlation.
For example, Au, Rh, Pd, and Ir all have similar excess electrons at 0 \vshe, and to a first order have similar barriers.
However, Au binds hydrogen weakly and deviates positively from the correlation, while Rh, Pd, and Ir bind hydrogen strongly and deviate negatively from the correlation.
A similar argument can be made regarding the relative deviations of Ag and Cu.
This shows an interplay between covalent and ionic interactions in driving the barrier energetics, with the ionic interactions providing the stronger interaction.

We can consider the primary driver of the differences in surface-charge density to be the (native) work function of each material, with secondary contributions from capacitances and solvation.~\cite{Kelly2022}
Indeed, we found a striking correlation between experimentally-tabulated work functions~\cite{Michaelson1977} and our excess-electron parameter (shown in the SI).
As a simplified, readily obtainable descriptor, we also plot in Figure~\ref{fig:material-correlations} the intrinsic barriers versus these experimentally-measured work functions for each metal.
We again see an excellent correlation.

This indicates that while the endstate energetics are driven by the formation of covalent bonds, the transition-state energetics are driven more strongly by the electrostatics.
Thus, different material properties can drive the behavior of these two states: covalent descriptors such as the d-band center may more strongly drive endstate binding energies, while the (native) material work function may be a larger driver of barrier energies due to their ionic nature.
Indeed, for many decades the experimental rates of reactions such as hydrogen evolution have been observed to correlate not only with binding energies, but also with material work functions~\cite{Bockris1947,Conway1957,Kita1966,Trasatti1972,Zeradjanin2017,Ostergaard2022,Shah2023}, with the physical reason remaining elusive~\cite{Ostergaard2022}.
Our relation between reaction barrier and work function offers a clear and compelling explanation for these experimental trends.
While surface electrostatic effects can affect all adsorbates,~\cite{Koper1999,Hyman2005,Karlberg2007,Chan2020,Ringe2023} we can generally expect that for reactions that create or destroy ions, electrostatic effects will be more significant at the transition state than the bound endstate (or vice versa in some systems such as \ce{CO2 -> $*$COO-}).
This offers a second degree of control in electrochemical systems.

In summary, in this work we show unique aspects of electrochemical barriers, as opposed to barriers at non-electrified interfaces.
First, the degree of charge transfer at the barrier drives many aspects of the barrier behavior, including its potential response, its adherence to the Hammond--Leffler postulate, and the curved Marcus-like behavior apparent over wide ranges in driving force.
Second, the interaction of the transition state with the specific electrode surface appears to be driven largely by electrostatic interactions with the surface, rather than the typical covalent descriptors involved in catalysis.
While we expect the degree of covalent versus ionic nature of each class of transition state to differ, we nonetheless expect that this identifies a different fundamental material driving force for electrochemical barriers.

\paragraph*{Methods.}
The solvated jellium method~\cite{Kastlunger2018} was used to perform electronic structure calculations.
Full details and parameters are available in the Supplementary Information (SI) which contains additional references.
\cite{Khorshidi2018, Zeng2022a, doi:10.1021/acs.jpcc.8b02465, Mortensen2005, Enkovaara2010, Larsen_2017, Schnur_2009, doi:10.1063/1.4948638, ase-paper, ISI:000175131400009}
Data is available from the corresponding author upon reasonable request.

\paragraph*{Acknowledgments.}
The authors acknowledge funding by the National Science Foundation under Award 1553365.
Calculations were performed at Brown's Center for Computation and Visualization.

\paragraph*{Author contributions.}
AP conceived of this study.
XC conducted the calculations with guidance from GK.
AP, XC, and GK developed the conclusions.
GK and AP developed the curved scaling model.
AP wrote the manuscript, and all authors contributed to the ideas within and its final form.

\paragraph*{Competing interests.}
The authors declare no competing interests.

\paragraph*{Materials \&\ correspondence.}
Direct inquiries to AP.

\bibliography{bibliography}
\end{document}

%% file: commands.tex
\newcommand{\barrierheight}{\ensuremath{E^\ddag}}
\newcommand{\barrierelectrons}{\ensuremath{N^\ddag}}
\newcommand{\reactionenergy}{\ensuremath{\Delta E}}
\newcommand{\pd}[3]{\ensuremath{\left( \frac{ \partial #1}{ \partial #2}\right)_{#3}}}
\newcommand{\pos}{\ensuremath{ \{ \vec{x}_i \} }}

\newcommand{\potential}{\ensuremath{\phi}}
\newcommand{\vshe}{\ensuremath{\mathrm{V}_{\mathrm{SHE}}}}